\documentclass[12pt]{vrchronology}

\usepackage{graphicx}
\usepackage{wrapfig}
\usepackage{amssymb}
\usepackage[english]{babel}
\usepackage{xurl}  % Allows long standards and product URLs to break cleanly.

\usepackage[bottom]{footmisc}

\usepackage[utf8]{inputenc}
\usepackage[T1]{fontenc}
\usepackage{lmodern}
\usepackage{enumitem}

\AtBeginDocument{}
\usepackage{pdflscape}

\usepackage{tikz}
\usepackage{float}
\floatstyle{plaintop}
\restylefloat{table}

\usepackage{placeins}
\usepackage{booktabs}
\usepackage{array}
\usepackage{tcolorbox}
\usepackage[hidelinks]{hyperref}
\setlist[itemize]{leftmargin=*,topsep=0.35\baselineskip,itemsep=0.15\baselineskip,parsep=0pt}
\setlist[enumerate]{leftmargin=*,topsep=0.35\baselineskip,itemsep=0.25\baselineskip,parsep=0pt}

\newcolumntype{L}[1]{>{\raggedright\arraybackslash}p{#1}}

\newtcolorbox{epigraphbox}{
  colback=white,
  colframe=black,
  boxrule=0.5pt,
  arc=0pt,
  outer arc=0pt,
  boxsep=0pt,
  left=8pt,
  right=8pt,
  top=6pt,
  bottom=6pt,
  before skip=0.6\baselineskip,
  after skip=0.6\baselineskip,
  fontupper=\itshape
}

\usetikzlibrary{fit, arrows,backgrounds,patterns,shapes,shapes.multipart,positioning,calc,decorations.markings}

\tikzset{
    opera/.style={draw, rectangle, align=center, text width=6cm, fill=white, font=\scriptsize},
    typnode/.style={anchor=north west, text width=12cm, inner sep=0mm},
    data/.style={draw=gray, rectangle, fill=white, font=\scriptsize, inner sep=0.5mm},
    datum/.style={font=\scriptsize, rotate=90, inner sep=1mm},
}

\def\evento#1#2#3;{
        \coordinate (A) at (0,{(#1-1900)/5});
        \coordinate (B) at (3,{#2});%
        \coordinate[right=5mm of A] (Z);
        \coordinate[left=1cm of B] (Y);
        \draw (A) -- (Z);
        \draw[-|] (Y) -- (B) node[data, pos=0.55] {#1};
        \draw (Z) -- (Y);
        \node[right=3mm of B, typnode] (D) {#3};
}

\def\eventoto#1#2#3;{
        \coordinate (A) at (0,{(#1-1972)/5});
        \coordinate (B) at (3,{#2});%
        \coordinate[right=5mm of A] (Z);
        \coordinate[left=1cm of B] (Y);
        \draw (A) -- (Z);
        \draw[-|] (Y) -- (B) node[data, pos=0.55] {#1};
        \draw (Z) -- (Y);
        \node[right=3mm of B, typnode] (D) {#3};
}

\def\eventototo#1#2#3;{
        \coordinate (A) at (0,{(#1-1982)/5});
        \coordinate (B) at (3,{#2});%
        \coordinate[right=5mm of A] (Z);
        \coordinate[left=1cm of B] (Y);
        \draw (A) -- (Z);
        \draw[-|] (Y) -- (B) node[data, pos=0.55] {#1};
        \draw (Z) -- (Y);
        \node[right=3mm of B, typnode] (D) {#3};
}

\def\eventotototo#1#2#3;{
        \coordinate (A) at (0,{(#1-1991)/5});
        \coordinate (B) at (3,{#2});%
        \coordinate[right=5mm of A] (Z);
        \coordinate[left=1cm of B] (Y);
        \draw (A) -- (Z);
        \draw[-|] (Y) -- (B) node[data, pos=0.55] {#1};
        \draw (Z) -- (Y);
        \node[right=3mm of B, typnode] (D) {#3};
}

\def\eventototototo#1#2#3;{
        \coordinate (A) at (0,{(#1-2007)/5});
        \coordinate (B) at (3,{#2});%
        \coordinate[right=5mm of A] (Z);
        \coordinate[left=1cm of B] (Y);
        \draw (A) -- (Z);
        \draw[-|] (Y) -- (B) node[data, pos=0.55] {#1};
        \draw (Z) -- (Y);
        \node[right=3mm of B, typnode] (D) {#3};
}

\def\eventotototototo#1#2#3;{
        \coordinate (A) at (0,{(#1-1991)/5});
        \coordinate (B) at (3,{#2});%
        \coordinate[right=5mm of A] (Z);
        \coordinate[left=1cm of B] (Y);
        \draw (A) -- (Z);
        \draw[-|] (Y) -- (B) node[data, pos=0.55] {#1};
        \draw (Z) -- (Y);
        \node[right=3mm of B, typnode] (D) {#3};
}

\newcommand*{\timelinecite}[2]{%
  \unskip\nobreak\hspace{0.22em}%
  \textnormal{(ref.\,#2)}%
}

\newcommand*{\timelineaxis}[1]{
        \draw[|->, -latex, draw] (0,0) -- (0,#1);
        \draw[-, dashed] (0,-0.5) -- (0,0);
        \foreach \y in {0,1,...,#1}{\draw (-0.1,\y) -- (0.1,\y);}
}

\newcommand*{\eventcontinue}[4]{
        \coordinate (A) at (0,{(#1-#4)/5});
        \coordinate (B) at (2,{#2});%
        \coordinate[right=5mm of A] (Z);
        \coordinate[left=1cm of B] (Y);
        \draw (A) -- (Z);
        \draw[-|] (Y) -- (B) node[data, pos=0.55] {#1};
        \draw (Z) -- (Y);
        \node[right=3mm of B, typnode, text width=5.5cm] (D)
          {\begin{NoHyper}#3\end{NoHyper}};
}

\newcommand*{\timelinecoljoin}{\hspace{0.5cm}}

\journal{ArXiv}

\begin{document}

\begin{frontmatter}


\title{A brief chronology of virtual reality: from laboratory systems to ubiquitous spatial computing, 1916--2026}



%----------------------------------------------------------------------------------------
%	TITLE
%----------------------------------------------------------------------------------------

\author[ualr]{Aryabrata Basu\corref{cor1}\fnref{formerinst}}
\ead{abasu@ualr.edu}
\cortext[cor1]{Corresponding author}
\address[ualr]{School of Computer and Information Sciences, University of Arkansas at Little Rock, Little Rock, Arkansas, United States}
\fntext[formerinst]{Former affiliation: Emory University, Atlanta, Georgia, United States}

% Use a fixed date so later arXiv rebuilds do not change the displayed PDF.
\date{September 29, 2026}

\begin{abstract}
\textit{Virtual reality (VR) developed through the progressive integration of displays, tracking, computation, interaction, and content rather than through a single device lineage. This article first establishes a systems-oriented vocabulary for virtual environments, immersion and presence, sensory feedback, and interactivity. It then presents two complementary chronologies: a general history of consequential VR developments from 1916 through 25 September 2026, and a focused history of ubiquitous VR design from 1991 through the same cutoff date. The latter follows the transition from assembled laboratory systems to commodity sensing, mobile displays, standalone headsets, mixed-reality platforms, and AI-mediated spatial computing. Together, the chronologies show that ubiquity can no longer be measured by portability or component count alone. It depends on a coherent low-latency perception--action loop and on whether the surrounding platform is deployable, interoperable, inclusive, safe, and trustworthy.}
\end{abstract}


\end{frontmatter}

%----------------------------------------------------------------------------------------
%	ARTICLE CONTENTS
%----------------------------------------------------------------------------------------
\par
\begin{epigraphbox}
``Equipped with his five senses, man explores the universe around him and calls the adventure Science.''
\par\smallskip\noindent\hfill{\normalfont --- Edwin Powell Hubble, \textit{The Nature of Science}, 1954}
\end{epigraphbox}

\section{Introduction}
\label{sec:introduction}

\begin{figure}[th]
  \begin{center}
     \includegraphics[width=1\linewidth]{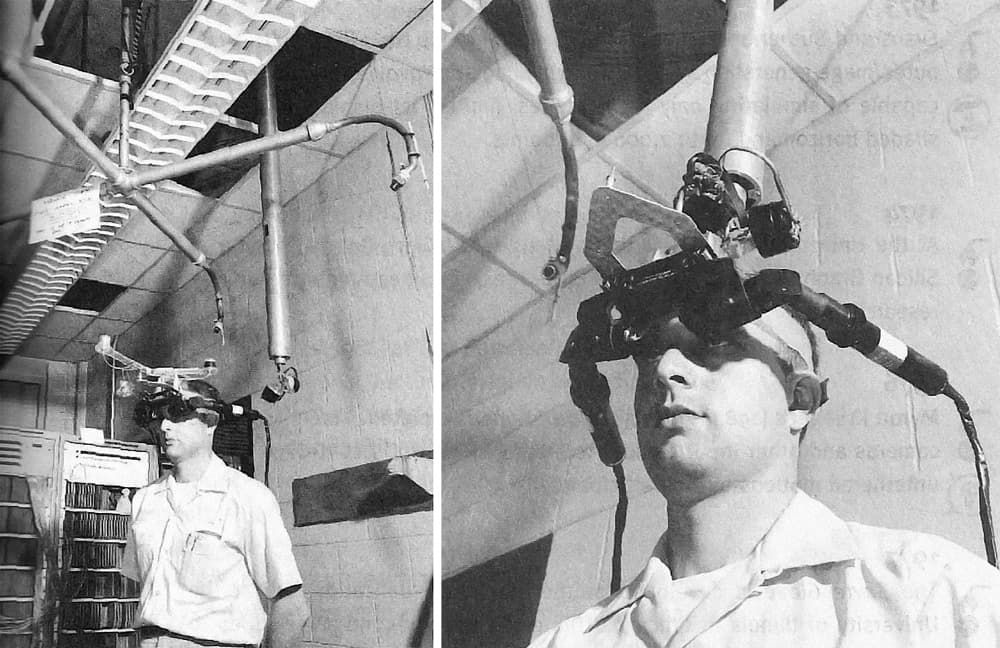}
  \end{center}
  \caption[The first head-mounted display]
    {Ivan Sutherland's head-mounted 3D display (c. 1968). The display was suspended from a counterbalanced mechanical arm and used ultrasonic transducers to track head movement. (Left) The system in use. (Right) Its principal components. Images reproduced from Sutherland (1968), with permission from Dr. Ivan Sutherland.}
  \label{fig:firsthmd}
\end{figure}

\textbf{Version note.} This substantially revised and expanded version of arXiv:1911.09605 retains its systems-oriented history of virtual reality and its ubiquitous-VR scope. It treats VR not as a single-device lineage but as the integration of stereoscopic display, head and body tracking, real-time graphics, spatial input, simulation, and multimodal feedback into a stable perception--action loop. That systems view explains why present-day strengths and limitations often echo earlier laboratory integration problems.

Ivan Sutherland's work provides a useful point of orientation. In 1963, Sketchpad combined an x-y vector display with a tracked light pen for computer-aided drawing \citep{DBLP:books/garland/Sutherland63}. Two years later, Sutherland described the ``ultimate display'' as ``a room within which the computer can control the existence of matter'' \citep{Sutherland1968}. Sutherland and Bob Sproull subsequently created a tracked, binocular head-mounted display that rendered imagery according to the user's changing head position and orientation. As Figure~\ref{fig:firsthmd} illustrates, the apparatus relied on a counterbalanced mechanical arm and ultrasonic tracking. It established the enduring objective of making a computational world respond coherently to the viewer's actions.

VR now spans simulation and training, industrial design, health, surgical planning, education, collaborative work, data visualization, and games. The field's history is therefore best read at two scales. The first chronology follows major technical and platform transitions across VR as a whole. The second asks a narrower systems question: when and how did VR become easier to acquire, transport, configure, share, maintain, and trust? Separating these views prevents consumer visibility from being mistaken for technical completeness and prevents hardware miniaturization from being mistaken for ubiquity.

The paper proceeds from concepts to evidence and then to synthesis. Section~\ref{sec:whatisvr} defines the experiential and systems elements used to interpret the record. Section~\ref{sec:historyofvr} presents the general chronology. Section~\ref{sec:ubiquitousvrsystem} then traces the distinct lineage of ubiquitous VR, followed by an assessment of the present trajectory and an open research agenda.

\FloatBarrier

\section{Conceptual framework for VR experiences}
\label{sec:whatisvr}

The enduring elements of a VR experience are a virtual environment, an immersive delivery system, sensory feedback, and interactivity. Contemporary systems make their dependencies clearer: immersion describes properties of the system, presence describes a user's response, and both emerge from a closed loop that includes sensing, state estimation, rendering, input, output, and application logic.

\subsection{Virtual environment}
\label{sec:vrworld}
A virtual environment (VE) is the content and subject matter of a virtual experience. It comprises virtual entities, their attributes, and the rules and relationships that govern their behavior. A VE ``capitalizes upon natural aspects of human perception by extending visual information in three spatial dimensions,'' may supplement this information with other stimuli and temporal change, and enables interaction with the displayed data \citep{wilson1999virtual}. VEs function as a communication medium for domains ranging from education, training, and entertainment to exploratory visualization, teleoperation, and experimental research.

\textit{Definition: A virtual environment describes a collection of objects in a virtual space and the rules and relationships governing those objects.}

\subsection{Immersion and presence}
\label{sec:immersion}
The terms \textit{immersion} and \textit{presence} are related but should not be used interchangeably. Following Slater \citep{slater2003note}, immersion is an objective property of the mediating system, whereas presence is the user's subjective response to that system:
\begin{itemize}[leftmargin=*]
\item \textit{Immersion is the objective level of sensory and interactive fidelity supplied by a VR system.}
\item \textit{Presence is a user's subjective psychological response to the mediated environment.}
\end{itemize}
Putting on an HMD can occlude the physical world and provide stereoscopic, head-coupled imagery, but the apparatus alone does not guarantee presence. Tracking fidelity, field of view, stereoscopy, audio, interaction, content, task, and individual differences all contribute. A meta-analysis found that technological immersion has a measurable but incomplete effect on presence, with field of view, stereoscopy, and tracking among the consequential features \citep{CummingsBailenson2016Presence}. Historically, device improvements expand the conditions under which presence may occur; they do not determine the experience by themselves.

\newpage
\subsection{Sensory feedback and tracking}
\label{sec:sensoryfeedback}
\begin{wrapfigure}[20]{r}{0.42\textwidth}
  \vspace{-0.75\baselineskip}
  \centering
  \includegraphics[width=\linewidth]{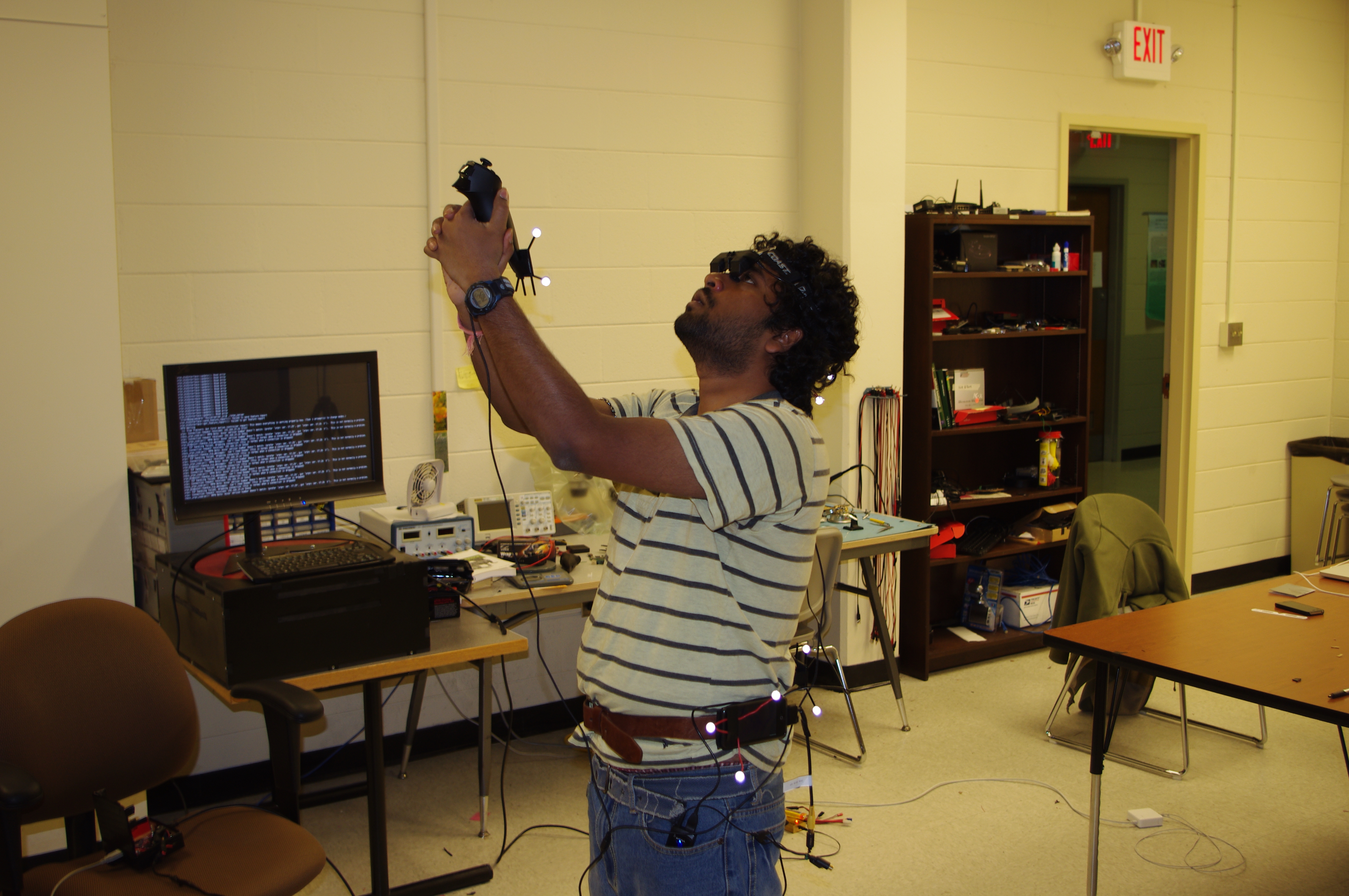}
  \caption[Sensory feedback by optical position tracking]
    {Optical position tracking in the former Virtual Experiences Lab at the University of Georgia: five OptiTrack Flex 3 cameras tracked retro-reflective markers at 100 FPS.}
  \label{fig:trackinginvel}
  \vspace{-0.5\baselineskip}
\end{wrapfigure}
VR allows a participant to act from an embodied perspective \citep{sherman2002understanding}. In a flight simulator, for example, the system estimates head motion and control input, updates the simulated aircraft and viewpoint, and returns visual, aural, and sometimes haptic consequences. Sensory feedback is therefore not output alone; it is the return path of a continuously updated loop. Early systems often depended on external optical or magnetic trackers such as the installation in Figure~\ref{fig:trackinginvel}. Contemporary headsets more commonly fuse cameras and inertial measurements for inside-out tracking and may also estimate hands, eyes, surfaces, and spatial anchors \citep{HanEtAl2022UmeTrack,HuEtAl2024SpatialTracking}.

Visual feedback remains predominant, but spatial audio, controller vibration, hand-contact cues, and physical props can materially shape an experience. Video passthrough also makes the relationship between sensed physical space and rendered content part of the feedback design. Errors or delays anywhere in this pipeline can break alignment, impair performance, or contribute to discomfort. Hardware is only part of the safety problem: a systematic review found that visual stimulation, locomotion, and exposure time also influence VR sickness \citep{SaredakisEtAl2020VRSickness}.

\textit{Definition: A VR system is an integrated collection of hardware, software, and content assembled to produce a VR experience.}

\textit{Definition: Position tracking is the sensing of the position and/or orientation of an object in the physical world.}

\subsection{Interactivity}
\label{sec:interactivity}
A VR experience becomes interactive when the user feedback loop \citep{sherman2002understanding} is closed: an action changes application state and the system returns an intelligible consequence. Locomotion and object manipulation are common examples, but interaction need not imitate physical reality. Controllers, tracked hands, gaze, voice, facial expression, and conventional inputs may be combined according to the task. Good interaction design concerns mappings, feedback, error recovery, and alternatives for different abilities, not simply the number of tracked input channels \citep{MottEtAl2020AccessibleByDesign,DudleyEtAl2023InclusiveImmersion}.

\subsection{A contemporary systems view}
The four experiential elements depend on a broader platform stack:
\begin{itemize}[leftmargin=*]
\item \textit{Sensing and state estimation} recover the motion of the head, hands, eyes, body, and surrounding space.
\item \textit{Scene and simulation models} define virtual objects, physical-world geometry, behavior, and shared state.
\item \textit{Rendering and feedback} transform that state into visual, aural, and haptic output under strict latency and power constraints.
\item \textit{Interaction} maps user intent to actions while providing discoverability, error recovery, and accessible alternatives.
\item \textit{Runtimes and standards} mediate application access to heterogeneous devices; OpenXR and WebXR are important contemporary examples \citep{Khronos2024OpenXR11,W3C2026WebXR}.
\item \textit{Data governance} determines what sensor data are retained, inferred, transmitted, and exposed to applications. Motion and spatial telemetry can reveal identity and behavior, so this layer is functional rather than merely administrative \citep{NairEtAl2023UniqueIdentification,MunillaGarridoEtAl2024PrivacySoK}.
\end{itemize}
This systems view supplies the vocabulary for interpreting both chronologies that follow. It also makes explicit that a VR experience is jointly produced by hardware, software, content, standards, and data practices.

\section{Historical chronology: systems, platforms, and convergence}
\label{sec:historyofvr}

\subsection{Scope, evidence, and reading order}
The chronology is selective but deliberately broad: an event is included when it materially changed display, sensing, interaction, real-time computation, content representation, platform access, or the evidence used to evaluate VR. The earlier panels emphasize foundational systems and platform formation; the finer-grained 2016--2026 panels emphasize the rapid convergence of standalone computing, mixed reality, standards, learned rendering, accessibility, and privacy. Dates refer to public demonstrations, shipment, or standards publication as specified in each entry. Peer-reviewed sources include a DOI in the bibliography whenever the publisher assigned one; product and standards milestones without DOIs are cited to primary records.

\subsection{Foundations and platform formation: 1916--2015}

The following five panels retain the original vertical-timeline mechanics and divide the foundational chronology into readable periods. Together they follow the movement from mechanical simulation and experimental displays to interactive graphics, tracked environments, projection systems, commodity sensing, and the first modern consumer-platform announcements.

\begin{table}[p]

\footnotesize   
    \begin{tikzpicture}[x=1cm,y=-7mm]
            \centering
       %draw horizontal line   
       \draw[|->, -latex, draw] (0,0) -- (0,25);
       \draw[-, dashed] (0,-0.5) -- (0,0);
       
       %draw years
        \foreach \y [evaluate=\y as \xear using int(1900+\y*5)] in {0,1,...,25}{ 
            \iffalse \draw (0,\y) node[left=2pt,anchor=east,xshift=0,font=\scriptsize] {$\xear$}; \fi
            \draw (-0.1,\y) -- (0.1,\y);
            }

        \evento{1916}{0}{U.S. Patent 1,183,492 was awarded to Albert B. Pratt for headgear carrying a gun and a sight aligned with the wearer's eyes. \citep{Pratt1916Patent}};
        \evento{1929}{2}{Edward Link developed the Link Trainer, an electromechanical flight simulator that placed pilots in a replicated cockpit with working instruments. \citep{USAFMuseumLinkTrainer}};
        \evento{1946}{5}{ENIAC, an early general-purpose electronic digital computer, was publicly announced at the University of Pennsylvania. \citep{Penn1946ENIAC}};
        \evento{1957}{7}{Morton Heilig invented the Sensorama, a noninteractive multisensory simulator combining stereoscopic film with sound, scent, wind, and vibration. \citep{HMHArchiveSensorama,Heilig1962SensoramaPatent}};
        \evento{1960}{9}{U.S. Patent 2,955,156 was granted to Morton Heilig for a stereoscopic television apparatus resembling a head-mounted display. \citep{Heilig1960TelespherePatent}};
        \evento{1961}{11}{Philco engineers Charles Comeau and James Bryan described Headsight, an HMD that coupled the wearer's head motion to a remote video camera---an early telepresence system. \citep{ComeauBryan1961Headsight}};
        \evento{1963}{13}{Ivan Sutherland introduced Sketchpad, an interactive computer-graphics system supporting drawing, selection, and constraints through a light pen and keyboard. \citep{DBLP:books/garland/Sutherland63}};
        \evento{1964}{15}{IBM completed Design Augmented by Computer (DAC-1) for General Motors after a joint project begun in the late 1950s. \citep{CHM1964DAC1}};
        \evento{1965}{17}{Ivan Sutherland described the ``ultimate display,'' a computational environment in which users could interact with objects unconstrained by physical reality. \citep{sutherland1965ultimate}};
        \evento{1967}{19}{Fred Brooks's group at the University of North Carolina began Project GROPE to provide haptic display of molecular-force fields. \citep{BrooksEtAl1990GROPE}};
        \evento{1968}{22}{Ivan Sutherland published ``A Head-Mounted Three-Dimensional Display.'' The tracked stereoscopic HMD used separate optics for each eye and mechanical and ultrasonic tracking. \citep{Sutherland1968}};

    \end{tikzpicture}

\caption{Historical VR chronology: 1916--1968.}\label{table:historicaltimeline}\vskip -1.5ex

\end{table}

\begin{table}[p]
\footnotesize   
    \begin{tikzpicture}[x=1cm,y=-7mm]
            \centering
       %draw horizontal line   
       \draw[|->, -latex, draw] (0,0) -- (0,25);
       \draw[-, dashed] (0,-0.5) -- (0,0);
       
       %draw years
        \foreach \y [evaluate=\y as \xear using int(1972+\y*5)] in {0,1,...,25}{ 
           \iffalse  \draw (0,\y) node[left=2pt,anchor=east,xshift=0,font=\scriptsize] {$\xear$}; \fi
            \draw (-0.1,\y) -- (0.1,\y);
            }

        \eventoto{1972}{0}{Atari's \emph{Pong} brought real-time multiplayer interactive graphics to a mass audience. \citep{StrongMuseumPong}};
        
        \eventoto{1973}{2}{Evans \& Sutherland and Rediffusion began their NOVOVIEW commercial-airline visual-simulator project; KLM bought the first system. \citep{EvansSutherland1973Novoview}};
        
        \eventoto{1974}{4}{At the University of Utah, Jim Clark completed an interactive B-spline design system using a head-mounted display; he later founded Silicon Graphics. \citep{Clark1974BSpline}};

        \eventoto{1975}{6}{Myron Krueger publicly exhibited a two-way VIDEOPLACE installation, using cameras and projection to place participants' silhouettes in a shared responsive environment. \citep{KruegerEtAl1985Videoplace}};

        \eventoto{1977}{10}{The Sayre Glove was developed at the University of Illinois at Chicago. Flexible light-conductive tubes estimated finger bending and hand configuration. \citep{EVL1977SayreGlove}};

        \eventoto{1979}{14}{Eric Howlett developed the Large Expanse Extra Perspective (LEEP) panoramic stereoscopic optical system, later adapted for wide-field HMDs. \citep{Howlett1979LEEP}};

    \end{tikzpicture}
\caption{Historical VR chronology: 1972--1979.}\vskip -1.5ex

\end{table}

\begin{table}[p]
\footnotesize   
    \begin{tikzpicture}[x=1cm,y=-7mm]
            \centering
       %draw horizontal line   
       \draw[|->, -latex, draw] (0,0) -- (0,25);
       \draw[-, dashed] (0,-0.5) -- (0,0);
       
       %draw years
        \foreach \y [evaluate=\y as \xear using int(1982+\y*5)] in {0,1,...,25}{ 
            \iffalse \draw (0,\y) node[left=2pt,anchor=east,xshift=0,font=\scriptsize] {$\xear$}; \fi
            \draw (-0.1,\y) -- (0.1,\y);
            }

        \eventototo{1982}{0}{Sara Bly mapped multivariate data to auditory parameters, advancing sonification as an interface technique. Thomas Furness's VCASS prototype used an HMD to present pilots with off-boresight target and flight-path information. \citep{Bly1982Sound,kocian1977visually}};

        \eventototo{1983}{5}{At MIT, Mark Callahan completed a personalized-computing HMD, while Christopher Schmandt published a stereoscopic workstation that registered spatial input with displayed imagery. \citep{Callahan1983Headset,Schmandt1983Workstation}};
        
        \eventototo{1984}{8.5}{Jaron Lanier and colleagues founded VPL Research, which commercialized the DataGlove, EyePhone, and networked RB2 system and helped popularize the term virtual reality. \citep{HCIMuseumVPL}};

        \eventototo{1985}{11.5}{Scott Fisher joined NASA Ames and directed the VIEW project. Michael McGreevy's low-cost Virtual Visual Environment Display (VIVED) became part of the broader VIEW workstation; VIEW was derived from VIVED, not its predecessor. \citep{Fisher2016VIEW,NASAViewCollection}};

        \eventototo{1987}{15}{VPL's DataGlove instrumented hand posture for virtual interaction, while Polhemus introduced the Isotrak electromagnetic six-degree-of-freedom tracker. \citep{HCIMuseumVPL,HCIMuseumIsotrak}};

        \eventototo{1989}{19}{VPL released the EyePhone and promoted its RB2 system; Mattel released the Power Glove for the Nintendo Entertainment System; and Fakespace began building the mechanically tracked BOOM display. \citep{HCIMuseumVPL,NVMPowerGlove,HCIMuseumBOOM}};

    \end{tikzpicture}

\caption{Historical VR chronology: 1982--1989.}\vskip -1.5ex

\end{table}

\begin{table}[p]
\footnotesize   
    \begin{tikzpicture}[x=1cm,y=-7mm]
            \centering
       %draw horizontal line   
       \draw[|->, -latex, draw] (0,0) -- (0,25);
       \draw[-, dashed] (0,-0.5) -- (0,0);
       
       %draw years
        \foreach \y [evaluate=\y as \xear using int(1991+\y*5)] in {0,1,...,25}{ 
            \iffalse \draw (0,\y) node[left=2pt,anchor=east,xshift=0,font=\scriptsize] {$\xear$}; \fi
            \draw (-0.1,\y) -- (0.1,\y);
            }

       \eventotototo{1991}{0}{Virtual Research Systems released the Virtual Research Flight Helmet, a US\$6,000 wide-field commercial HMD. \citep{DeutschesMuseum1991FlightHelmet}};

       \eventotototo{1992}{2}{The CAVE was presented as a projection-based alternative to HMDs. NASA's Virtual Wind Tunnel used a BOOM and DataGlove to explore aerodynamic flow fields. \citep{DBLP:journals/cacm/Cruz-NeiraSDKH92,BrysonLevit1992VirtualWindTunnel}};

       \eventotototo{1993}{6}{The IEEE Symposium on Research Frontiers in Virtual Reality (VRAIS '93) established a conference lineage later consolidated as IEEE VR. Thomas Massie's MIT work produced the PHANToM desktop force-feedback interface. \citep{IEEE1993VRAIS,MassieSalisbury1994PHANToM}};

       \eventotototo{1994}{9}{The VROOM venue at SIGGRAPH demonstrated more than 40 applications running in CAVE systems. \citep{SIGGRAPH1994VROOM}};

       \eventotototo{1995}{12}{Virtual I-O released the i-glasses! personal display, packaging stereoscopic viewing and optional head tracking for the consumer market. \citep{GoogleArts1995IGlasses}};

       \eventotototo{1996}{16}{Ascension Technology introduced MotionStar, a wireless magnetic motion-capture system that reduced cabling for multi-receiver body tracking. \citep{Ascension1996MotionStar}};

       \eventotototo{1998}{19}{Disney opened DisneyQuest with HMD- and projection-based attractions. Six-surface CAVE-style systems were commissioned at KTH in Sweden and at Gifu's VR Techno Center in Japan; the record does not support an unqualified single ``first.'' \citep{DisneyQuest1998,Mine2003DisneyVR,KTH1998VRCube,YamadaEtAl1999COSMOS}};

       \eventotototo{1999}{22}{Hirokazu Kato and Mark Billinghurst published the marker-tracking and HMD-calibration work underlying ARToolKit, which made video-based fiducial tracking widely accessible. \citep{KatoBillinghurst1999ARToolKit}};

       \eventotototo{2000}{25}{Iowa State University opened C6, the first six-sided immersive room in the United States. \citep{IowaState2000C6}};

    \end{tikzpicture}

\caption{Historical VR chronology: 1991--2000.}\vskip -1.5ex

\end{table}

\begin{table}[!t]
    \footnotesize   
    \begin{tikzpicture}[x=1cm,y=-5.5mm]
            \centering
       %draw horizontal line   
       \draw[|->, -latex, draw] (0,0) -- (0,15);
       \draw[-, dashed] (0,-0.5) -- (0,0);
       
       %draw years
        \foreach \y [evaluate=\y as \xear using int(2007+\y*5)] in {0,1,...,15}{ 
            \iffalse \draw (0,\y) node[left=2pt,anchor=east,xshift=0,font=\scriptsize] {$\xear$}; \fi
            \draw (-0.1,\y) -- (0.1,\y);
            }

       \eventototototo{2007}{0}{Google published the first Street View imagery, making navigable street-level panoramas available on the web. \citep{Google2007StreetView}};

       \eventototototo{2011}{3}{Basu et al. began the UCAVE line of portable, smartphone-based immersive systems for rapidly deployable virtual collaboration. \citep{DBLP:conf/cscw/BasuRJ12}};

       \eventototototo{2012}{5.2}{USC's MxR Lab introduced FOV2GO, an open viewer and software kit for smartphone VR. Palmer Luckey also launched the Oculus Rift Kickstarter campaign. \citep{USCICT2012FOV2Go,Kickstarter2012OculusRift}};

       \eventototototo{2013}{9}{Valve's Michael Abrash publicly analyzed display persistence, judder, latency, and other barriers to perceptually stable consumer VR. \citep{Abrash2013VRHard}};
       
       \eventototototo{2014}{11.5}{Facebook agreed to acquire Oculus VR for approximately US\$2 billion. \citep{Facebook2014OculusAcquisition}};
       
       \eventototototo{2015}{13.2}{HTC and Valve announced the Vive headset, SteamVR tracking, and tracked controllers, establishing the platform later shipped in 2016. \citep{HTC2015ViveAnnouncement}};

    \end{tikzpicture}

\caption{Historical VR chronology: 2007--2015.}\vskip -1.5ex

\end{table}

\FloatBarrier
\subsection{From laboratory systems to mobile platforms}

The five foundational panels show integration rather than a sequence of isolated inventions. The span from Sutherland's tracked HMD to application-oriented systems such as VCASS reflects the slow convergence of reliable tracking, displays, and real-time graphics. By the late 2000s, smartphones combined displays, inertial sensors, processors, and software distribution at mass-market scale. They made portable VR easier to reproduce and extended a democratizing lineage that includes Randolph Pausch's 1991 ``Five Dollar a Day'' system \citep{Pausch:1991:VRF:108844.108913}.

In 2011, Basu et al. built an early untethered research system for portable, rapid deployment using a smartphone display and internal inertial sensing \citep{DBLP:conf/cscw/BasuRJ12}. MuVR pursued low-cost multi-user operation \citep{DBLP:conf/vr/ThomasBGS14}, while Bachmann et al. combined body-worn inertial sensors, zero-velocity updates, outdoor GPS, and obstacle sensing \citep{DBLP:conf/vr/HodgsonBWBO12}. Together, these projects represent a broader shift from bespoke laboratory installations toward reproducible mobile systems rather than a single first.

Across these eras, the invariant is not a refresh-rate threshold but a closed perception--action loop: the system must sense the user, update its state, render the consequence, and return coherent feedback with sufficient stability and low latency. Display rate is only one component; tracking quality, motion-to-photon latency, prediction, content motion, locomotion design, and user characteristics also affect comfort and performance \citep{SaredakisEtAl2020VRSickness}.

\subsection{Contemporary convergence: 2016--2026}
\label{sec:extendedchronology}

Table~\ref{table:extendedtimeline} presents eight finer-grained panels, each pair read down the left and then the right. Entries trace changes in deployment, sensing, interaction, rendering, portability, safety, privacy, and governance. Dates distinguish announcements from availability and flag products not shipping by the 25 September 2026 cutoff. Peer-reviewed sources include assigned DOIs; other entries cite first-party records.

\clearpage
\begin{landscape}
\begin{table}[p]
\centering
\footnotesize
\begin{minipage}[t]{0.485\textwidth}
\vspace{0pt}
\centering
\textbf{2016}\par\smallskip
\begin{tikzpicture}[x=1cm,y=-5mm,typnode/.append style={text width=5.9cm,font=\scriptsize}]
    \timelineaxis{22}
    \eventcontinue{2016}{0}{A meta-analysis of 83 studies separated technological immersion from subjective presence and quantified how tracking, stereoscopy, and field of view relate to presence. \timelinecite{CummingsBailenson2016Presence}{22}}{2016}
    \eventcontinue{2016}{4}{\textbf{28 March:} The consumer Oculus Rift began shipping with positional tracking and integrated audio, turning the developer-kit revival into a supported \mbox{PC VR platform.\ \timelinecite{Meta2016Rift}{87}}}{2016}
    \eventcontinue{2016}{7.5}{\textbf{30 March:} Microsoft began shipping the self-contained HoloLens Development Edition with spatial mapping, gaze, gesture, and voice interaction. \timelinecite{Microsoft2016HoloLensShips}{95}}{2016}
    \eventcontinue{2016}{10.5}{\textbf{5 April:} HTC Vive shipments began; Lighthouse base stations and tracked controllers packaged room-scale six-degree-of-freedom interaction for consumers. \timelinecite{HTC2016Vive}{54}}{2016}
    \eventcontinue{2016}{14}{\textbf{October:} \emph{Holoportation} demonstrated real-time capture, transmission, and reconstruction of a remote person for mixed-reality telepresence. \timelinecite{OrtsEscolanoEtAl2016Holoportation}{107}}{2016}
    \eventcontinue{2016}{18}{\textbf{6 December:} Oculus Touch shipped as paired tracked controllers, establishing two-handed controller-mediated ``hand presence'' as a mainstream VR interaction pattern. \timelinecite{Meta2016Touch}{82}}{2016}
\end{tikzpicture}
\end{minipage}\timelinecoljoin
\begin{minipage}[t]{0.485\textwidth}
\vspace{0pt}
\centering
\textbf{2017--2018}\par\smallskip
\begin{tikzpicture}[x=1cm,y=-5mm,typnode/.append style={text width=5.9cm,font=\fontsize{6.5}{7.2}\selectfont}]
    \timelineaxis{23}
    \eventcontinue{2017}{0}{\textbf{27 February:} Khronos announced the OpenXR working group to create a royalty-free interface between XR applications and runtimes. \timelinecite{Khronos2017OpenXR}{63}}{2017}
    \eventcontinue{2017}{2.5}{\textbf{May:} \emph{ShareVR} enabled one HMD user and several non-HMD users to share a co-located experience through projection, tracked props, and asymmetric roles. \timelinecite{GugenheimerEtAl2017ShareVR}{40}}{2017}
    \eventcontinue{2017}{5}{\textbf{17 May:} Google announced standalone Daydream headsets and WorldSense inside-out tracking, removing the phone slot, cable, and external tracking infrastructure. \timelinecite{Google2017StandaloneDaydream}{32}}{2017}
    \eventcontinue{2017}{7.5}{\textbf{July:} \emph{WiCapture} demonstrated sub-centimeter three-dimensional tracking with commodity Wi-Fi signals. \timelinecite{KotaruKatti2017WiCapture}{70}}{2017}
    \eventcontinue{2017}{10}{A perceptual study quantified spatial and temporal limits for foveated rendering, including the effect of gaze-to-display latency. \timelinecite{AlbertEtAl2017FoveatedRendering}{3}}{2017}
    \eventcontinue{2018}{12.5}{\textbf{April:} \emph{FaceDisplay} placed touch displays and cameras outside an HMD so nearby people could see and manipulate its virtual scene. \timelinecite{GugenheimerEtAl2018FaceDisplay}{41}}{2017}
    \eventcontinue{2018}{15}{\textbf{1 May:} Oculus Go became available as a self-contained US\$199 headset, consolidating display, compute, tracking, and distribution into an inexpensive appliance. \timelinecite{Meta2018OculusGo}{83}}{2017}
    \eventcontinue{2018}{17}{\textbf{May:} Lenovo Mirage Solo brought WorldSense six-degree-of-freedom inside-out tracking to a commercial standalone Daydream headset. \timelinecite{Google2018MirageSolo}{34}}{2017}
    \eventcontinue{2018}{19}{\textbf{June:} Foveated streaming linked eye tracking to network efficiency by assigning panoramic-video quality according to gaze. \timelinecite{RomeroRondonEtAl2018FoveatedStreaming}{114}}{2017}
    \eventcontinue{2018}{21}{\textbf{August:} Khronos demonstrated early OpenXR prototypes across vendors, showing that one application interface could target incompatible XR systems. \timelinecite{Khronos2018OpenXRDemos}{64}}{2017}
\end{tikzpicture}
\end{minipage}
\caption{Extended VR/XR timeline: 2016--2018. Continue from the bottom of the left panel to the top of the right panel.}
\label{table:extendedtimeline}
\end{table}

\begin{table}[p]
\centering
\footnotesize
\begin{minipage}[t]{0.485\textwidth}
\vspace{0pt}
\centering
\textbf{2019--2020}\par\smallskip
\begin{tikzpicture}[x=1cm,y=-4.8mm,typnode/.append style={text width=5.9cm,font=\fontsize{6.5}{7.2}\selectfont}]
    \timelineaxis{24}
    \eventcontinue{2019}{0}{\textbf{24 February:} Microsoft announced HoloLens 2 with a wider field of view, eye and articulated-hand tracking, depth sensing, and direct hologram manipulation. \timelinecite{Microsoft2019HoloLens2}{96}}{2019}
    \eventcontinue{2019}{2.5}{\textbf{18 March:} Khronos released the provisional OpenXR 0.90 specification for public implementation and review. \timelinecite{Khronos2019OpenXR090}{65}}{2019}
    \eventcontinue{2019}{5}{\textbf{21 May:} Oculus Quest and Rift S shipped with camera-based Insight tracking; Quest paired six-degree-of-freedom tracking with standalone compute. \timelinecite{Meta2019QuestRiftS}{84}}{2019}
    \eventcontinue{2019}{7.5}{\textbf{29 July:} OpenXR 1.0 was ratified, creating a stable application/runtime boundary and a shared portability target. \timelinecite{Khronos2019OpenXR10}{66}}{2019}
    \eventcontinue{2019}{10}{\textbf{November:} \emph{DeepFovea} reconstructed peripheral imagery from sparse samples with a generative network, reducing foveated-rendering cost. \timelinecite{KaplanyanEtAl2019DeepFovea}{60}}{2019}
    \eventcontinue{2019}{12.5}{\textbf{December:} Quest optical hand tracking entered public beta, turning controller-free input into a software-delivered platform feature. \timelinecite{Meta2019QuestHandTracking}{85}}{2019}
    \eventcontinue{2020}{15}{\textbf{April:} A review of 38 higher-education studies exposed recurring weaknesses in learning theory, experimental design, and evaluation. \timelinecite{RadiantiEtAl2020HigherEducation}{113}}{2019}
    \eventcontinue{2020}{17.5}{\textbf{28 July:} The OpenXR Adopter Program introduced conformance testing and announced conformant Microsoft and Oculus implementations. \timelinecite{Khronos2020OpenXRConformance}{67}}{2019}
    \eventcontinue{2020}{20}{\textbf{August:} Neural Radiance Fields synthesized photorealistic novel views from posed images and became foundational to neural immersive-content capture. \timelinecite{MildenhallEtAl2020NeRF}{97}}{2019}
    \eventcontinue{2020}{22.5}{\textbf{13 October:} Quest 2 shipped at US\$299 with greater resolution and processing performance, accelerating standalone six-degree-of-freedom VR. \timelinecite{Meta2020Quest2}{86}}{2019}
\end{tikzpicture}
\end{minipage}\timelinecoljoin
\begin{minipage}[t]{0.485\textwidth}
\vspace{0pt}
\centering
\textbf{2021--2022}\par\smallskip
\begin{tikzpicture}[x=1cm,y=-4.8mm,typnode/.append style={text width=5.9cm,font=\scriptsize}]
    \timelineaxis{22}
    \eventcontinue{2021}{0}{\textbf{2 March:} Microsoft announced Mesh, a shared mixed-reality platform spanning HoloLens, VR headsets, phones, tablets, and PCs. \timelinecite{Microsoft2021Mesh}{75}}{2021}
    \eventcontinue{2021}{4}{\textbf{May:} A CHI study showed that head-and-hand motion can identify VR users, establishing behavioral telemetry as a security and privacy concern. \timelinecite{LiebersEtAl2021BehavioralBiometrics}{77}}{2021}
    \eventcontinue{2021}{8}{\textbf{19 August:} Horizon Workrooms entered open beta with avatars, spatial audio, hand tracking, virtual screens, and mixed-reality desk features. \timelinecite{Meta2021Workrooms}{88}}{2021}
    \eventcontinue{2022}{12}{\textbf{31 March:} W3C published a Candidate Recommendation Snapshot of WebXR, standardizing browser access to immersive devices and sensors. \timelinecite{W3C2022WebXR}{139}}{2021}
    \eventcontinue{2022}{16}{\textbf{25 October:} Quest Pro shipped with color stereoscopic passthrough, pancake optics, eye/face tracking, and self-tracking controllers. \timelinecite{Meta2022QuestPro}{89}}{2021}
    \eventcontinue{2022}{20}{\textbf{December:} \emph{UmeTrack} integrated multi-view egocentric hand detection, tracking, and model fitting for robust articulated input. \timelinecite{HanEtAl2022UmeTrack}{42}}{2021}
\end{tikzpicture}
\end{minipage}
\caption{Extended VR/XR timeline: 2019--2022. Continue from the bottom of the left panel to the top of the right panel.}
\end{table}

\begin{table}[p]
\centering
\footnotesize
\begin{minipage}[t]{0.485\textwidth}
\vspace{0pt}
\centering
\textbf{2023}\par\smallskip
\begin{tikzpicture}[x=1cm,y=-5mm,typnode/.append style={text width=5.9cm,font=\scriptsize}]
    \timelineaxis{22}
    \eventcontinue{2023}{0}{\textbf{5 June:} Apple announced Vision Pro and the visionOS ``spatial computer'' model centered on video passthrough and eye, hand, and voice input. \timelinecite{Apple2023VisionPro}{4}}{2023}
    \eventcontinue{2023}{5}{\textbf{July:} 3D Gaussian Splatting achieved high-quality novel-view rendering at real-time rates using explicit three-dimensional Gaussians. \timelinecite{KerblEtAl2023GaussianSplatting}{62}}{2023}
    \eventcontinue{2023}{10}{\textbf{July:} Attention-aware foveated rendering combined gaze with visual-attention modeling to account for conspicuous peripheral content. \timelinecite{KrajancichEtAl2023AttentionAware}{71}}{2023}
    \eventcontinue{2023}{15}{\textbf{August:} Motion from 55,541 VR users enabled accurate re-identification, showing that head-and-hand trajectories can act as biometric data. \timelinecite{NairEtAl2023UniqueIdentification}{101}}{2023}
    \eventcontinue{2023}{20}{\textbf{10 October:} Quest 3 shipped with full-color passthrough, a depth sensor, and mixed-reality scene understanding in a mainstream standalone headset. \timelinecite{Meta2023Quest3}{90}}{2023}
\end{tikzpicture}
\end{minipage}\timelinecoljoin
\begin{minipage}[t]{0.485\textwidth}
\vspace{0pt}
\centering
\textbf{2024}\par\smallskip
\begin{tikzpicture}[x=1cm,y=-5mm,typnode/.append style={text width=5.9cm,font=\scriptsize}]
    \timelineaxis{22}
    \eventcontinue{2024}{0}{\textbf{2 February:} Apple Vision Pro shipped in the United States with eye-and-hand input, high-resolution passthrough, spatial video, and desktop-class applications. \timelinecite{Apple2024VisionProShips}{6}}{2024}
    \eventcontinue{2024}{5}{A systematization of 74 publications organized VR privacy threats, data attributes, adversaries, and countermeasures. \timelinecite{MunillaGarridoEtAl2024PrivacySoK}{100}}{2024}
    \eventcontinue{2024}{10}{\textbf{15 April:} OpenXR 1.1 moved widely used extensions---including local-floor spaces, quad-view foveation, and palm pose---into the core specification. \timelinecite{Khronos2024OpenXR11}{68}}{2024}
    \eventcontinue{2024}{15}{\textbf{25 September:} Meta revealed the Orion AR-glasses prototype with transparent displays, contextual AI, eye/hand tracking, and an electromyography wristband. \timelinecite{Meta2024Orion}{91}}{2024}
    \eventcontinue{2024}{20}{\textbf{November:} A Vision Pro/Quest 3 comparison found sub-centimeter relative pose error in most cases but absolute errors above 10~cm under difficult conditions. \timelinecite{HuEtAl2024SpatialTracking}{57}}{2024}
\end{tikzpicture}
\end{minipage}
\caption{Extended VR/XR timeline: 2023--November 2024. Continue from the bottom of the left panel to the top of the right panel.}
\end{table}

\begin{table}[p]
\centering
\footnotesize
\begin{minipage}[t]{0.485\textwidth}
\vspace{0pt}
\centering
\textbf{December 2024--2025}\par\smallskip
\begin{tikzpicture}[x=1cm,y=-5mm,typnode/.append style={text width=5.9cm,font=\scriptsize}]
    \timelineaxis{20}
    \eventcontinue{2024}{0}{\textbf{12 December:} Google announced Android XR with Samsung and Qualcomm as a headset-and-glasses platform integrated with Gemini. \timelinecite{Google2024AndroidXR}{35}}{2024}
    \eventcontinue{2025}{4.5}{\textbf{20 May:} Google demonstrated Android XR glasses with Gemini, cameras, microphones, speakers, and optional in-lens displays. \timelinecite{Google2025AndroidXRGlasses}{37}}{2024}
    \eventcontinue{2025}{9}{\textbf{17 September:} Meta announced Ray-Ban Display glasses and an EMG wristband; U.S. retail availability began 30 September. \timelinecite{Meta2025RayBanDisplay}{92}}{2024}
    \eventcontinue{2025}{13.5}{\textbf{15 October:} Apple announced an M5 Vision Pro with a redesigned Dual Knit Band, an incremental compute and comfort revision. \timelinecite{Apple2025VisionProM5}{7}}{2024}
    \eventcontinue{2025}{18}{\textbf{21 October:} Samsung introduced Galaxy XR, the first Android XR headset, with eye/hand/voice input, Gemini, OpenXR, WebXR, and Unity support. {117, 36}}{2024}
\end{tikzpicture}
\end{minipage}\timelinecoljoin
\begin{minipage}[t]{0.485\textwidth}
\vspace{0pt}
\centering
\textbf{2026 to 25 September}\par\smallskip
\begin{tikzpicture}[x=1cm,y=-5mm,typnode/.append style={text width=5.9cm,font=\scriptsize}]
    \timelineaxis{20}
    \eventcontinue{2026}{0}{\textbf{7 April:} Android XR added experimental 2D-to-3D spatialization, persistent app placement, visible physical hands, and tracking/accessibility improvements. \timelinecite{Google2026AndroidXRUpdate}{38}}{2026}
    \eventcontinue{2026}{5.5}{\textbf{19 May:} Google described audio-first and display Android XR eyewear. Audio glasses were slated for later autumn; no display-glasses date was given. \timelinecite{Google2026IntelligentEyewear}{39}}{2026}
    \eventcontinue{2026}{11}{\textbf{9 June:} W3C issued a new WebXR Candidate Recommendation Draft while retaining the March 2022 Candidate Recommendation Snapshot. \timelinecite{W3C2026WebXR}{140}}{2026}
    \eventcontinue{2026}{16.5}{\textbf{23 September:} Meta announced roughly 100~g split-compute VR Glasses with a 5K micro-OLED display and eye/hand/voice input for spring 2027; the product had not shipped by the cutoff. \timelinecite{Meta2026VRGlasses}{93}}{2026}
\end{tikzpicture}
\end{minipage}
\caption{Extended VR/XR timeline: December 2024--25 September 2026. Continue from the bottom of the left panel to the top of the right panel.}
\end{table}
\end{landscape}

\FloatBarrier

\subsection{Synthesis: integration shifts responsibility to the platform}
The 2016--2026 period extends the paper's original democratization argument but also complicates it. External computers, base stations, and controllers increasingly gave way to onboard visual--inertial tracking, optical hand tracking, and standalone compute. Perceptual rendering and learned scene representations shifted part of the performance problem from brute-force graphics toward models of gaze, attention, and scene appearance. OpenXR and WebXR reduced application fragmentation, while video passthrough made mixed reality a default operating mode rather than a specialized capability. At the same time, the sensors that make systems easier to deploy---cameras, microphones, eye tracking, hand tracking, spatial maps, and continuous motion telemetry---produce intimate data. Ubiquity therefore no longer means portability alone: it also depends on interoperability, accessibility, reliable sensing, transparent AI behavior, and privacy-preserving data governance.

\section{From virtual reality to ubiquitous VR}
\label{sec:ubiquitousvrsystem}

The general chronology describes what VR systems made possible. This section applies a narrower lens: when did those capabilities become practical to acquire, reproduce, transport, configure, share, maintain, and trust? The distinction matters because a technically advanced system may remain far from ubiquitous if it depends on specialized infrastructure, inaccessible interaction, incompatible software, or opaque sensing.

The vision of ubiquitous computing in Mark Weiser's words \citep{weiser1994creating} is that ``a good tool is an invisible tool. By invisible, I mean that the tool does not intrude on your consciousness; you focus on the task, not the tool.'' Fewer wearable components can reduce setup burden, but component count alone is no longer an adequate measure of invisibility. A standalone headset may hide external computers and base stations while adding cameras, microphones, eye tracking, hand tracking, spatial maps, accounts, network services, and calibration procedures. Ubiquitous VR should therefore be evaluated by the total friction of deployment: cost, setup time, tracking robustness, portability, application compatibility, accessibility, safety, maintenance, and the governance of sensed data. In 1991, Pausch argued that the field would benefit from putting VR technology into as many researchers' hands as possible \citep{Pausch:1991:VRF:108844.108913}. That democratizing goal remains relevant, but access now includes the ability to use, understand, trust, and modify a system---not merely to acquire it.

\subsection{Design lineage: reducing deployment friction}
\label{sec:ubiquitousvrsystemdesignevolution}

Motion-based games and natural-interface research produced a wide variety of accessible, off-the-shelf sensing and input devices. These components reduced barriers to immersive-system development, lowered costs, and increased the reproducibility of VR research \citep{lee2008hacking,wingrave2010wiimote}. The design lineage begins, however, with Randy Pausch's explicit 1991 effort to construct a complete low-cost system \citep{Pausch:1991:VRF:108844.108913}.

Pausch's ``Five Dollar a Day'' VR system used an 80386-based IBM PC, a Polhemus 3Space Isotrak, two Reflection Technology Private Eye displays, and a Mattel Power Glove. At the time, the complete system cost less than US\$5,000. Its displays rendered monochrome wireframe objects at $720 \times 280$ pixels. Pausch emphasized a coherent, low-latency experience over high-resolution graphics or stereoscopy \citep{Pausch:1991:VRF:108844.108913}. The work demonstrated the value of redesigning systems so that VR could be studied more widely.

Designing a broadly deployable VR platform requires evaluating the display, input mappings, tracking, software, and their integration. Pausch compared head-tracked and non-head-tracked camera control in a visual-search task and found that head tracking reduced task-completion time by helping participants build a better internal representation of the environment \citep{pausch1993user}.

Building on Pausch's early work, we introduced the Ubiquitous Collaborative Activity Virtual Environment (UCAVE) framework in 2011 \citep{DBLP:conf/cscw/BasuRJ12}. UCAVEs are portable immersive environments that use mobile communication platforms, motion trackers, and displays to support ad hoc virtual collaboration.

Following the UCAVE framework, Anthony Steed described the design and implementation of an iPhone-based HMD system in 2013 \citep{steed2013design}. A key design choice used the phone itself as an unseen touch controller. The principal implementation challenge was alignment between the phone's inertial sensor and a Freespace head tracker: without an external reference, their yaw estimates rapidly diverged. Steed introduced a gesture for approximate realignment and a clutch for precise realignment. The system rendered virtual environments containing a few thousand polygons at 60~Hz with approximately 100~ms latency.

Our subsequent work addressed registration between different inertial sensors by introducing a portable electromagnetic tracker for six-degree-of-freedom single-hand input \citep{6184191}. Integration with UCAVE provided a common coordinate frame in an untethered configuration. The system did not track the user's translational position; this was sufficient for its orientation-centered use case, although room-scale systems later made six-degree-of-freedom head tracking a common expectation.

Contemporaneous work examined combinations of consumer gaming hardware, including the Razer Hydra, Wiimote, and Microsoft Kinect \citep{lu1virtual}. It showed that three-dimensional motion tracking could benefit from complementary analog controls and buttons.

Thomas et al. subsequently presented MuVR, a multi-user VR platform \citep{DBLP:conf/vr/ThomasBGS14}. They identified limited access, dependence on external rendering and tracking infrastructure, and setup time as barriers to ubiquitous deployment. MuVR used an Oculus Rift DK1 and a smartphone worn at the hips in a modular configuration related to the UCAVE prototype \citep{6184191}. Its contribution was to treat deployment logistics and simultaneous users as system-design requirements rather than afterthoughts.

In 2015, Ponto et al. introduced DSCVR, a commodity hybrid VR system \citep{ponto2015dscvr}. The work documented design considerations and observations from building a CAVE-like installation \citep{DBLP:journals/cacm/Cruz-NeiraSDKH92} with commodity technology. Although it was not a mobile system, it pursued the same democratizing goal by reducing the cost and specialization required for a room-scale installation.

\subsection{Ubiquitous VR chronology: 1991--2026}

The timeline below treats ``ubiquitous'' as a systems property rather than a synonym for a small headset. Milestones are included when they materially widened public access, reduced cost or setup, enabled portable or standalone use, supported shared deployment, turned consumer sensing into a reusable research component, or improved application portability, accessibility, and trust. This scope includes consequential failures and research prototypes as well as successful products because each exposed a deployment constraint. Geographic labels identify the originating organization or ecosystem rather than the only place of use. Academic works cite a publisher DOI whenever one was assigned; dated product and standards records generally have no DOI and are cited to the responsible organization's primary record. Read down the left panel and then continue from the top of the right panel across four landscape spreads through the cutoff date of 25 September 2026.

\begin{table}[H]
\centering
\small
\caption{Analytical lens for the ubiquitous VR chronology.}
\begin{tabular}{@{}L{0.24\textwidth}L{0.68\textwidth}@{}}
\toprule
Dimension & Inclusion signal \\
\midrule
Access and cost & Lowers acquisition cost or moves immersive capability beyond specialist laboratories. \\
Setup and mobility & Reduces calibration, cabling, external infrastructure, or time required to deploy and recover a system. \\
Sensing and interaction & Makes spatial input more robust, natural, reusable, or available through commodity components. \\
Participation and scale & Supports shared, public, collaborative, educational, enterprise, or multi-user deployment. \\
Portability and lifecycle & Improves application interoperability, distribution, maintenance, preservation, or ecosystem continuity. \\
Human and institutional fit & Advances accessibility, comfort, safety, privacy, understandable sensing, or accountable platform behavior. \\
Evidence and reach & Establishes consequential research evidence, documents a failure mode, or broadens the geographic availability of VR. \\
\bottomrule
\end{tabular}
\label{table:ubiquitylens}
\end{table}

\clearpage
\begin{landscape}
\begin{table}[p]
\centering
\footnotesize
\begin{minipage}[t]{0.485\textwidth}
\vspace{0pt}\centering
\textbf{1991--2006: lowering the laboratory barrier}\par\smallskip
\begin{tikzpicture}[x=1cm,y=-4.7mm,typnode/.append style={text width=5.9cm,font=\fontsize{6.5}{7.2}\selectfont}]
    \timelineaxis{24}
    \eventcontinue{1991}{0}{\textbf{United States:} Pausch assembled a complete low-cost system from commodity components and argued for putting VR into many more researchers' hands. \timelinecite{Pausch:1991:VRF:108844.108913}{108}}{1991}
    \eventcontinue{1991}{3.1}{\textbf{United Kingdom:} Virtuality brought networked, tracked HMD play into public arcades, making immersive multiplayer VR a paid public experience rather than a laboratory demonstration. \timelinecite{HCIMuseum1991Virtuality}{43}}{1991}
    \eventcontinue{1993}{6.2}{\textbf{United States:} Pausch, Shackelford, and Proffitt compared head-tracked and stationary displays, placing deployable HMD choices under user-study scrutiny. \timelinecite{pausch1993user}{109}}{1991}
    \eventcontinue{1994}{9.3}{\textbf{Germany:} GMD's Responsive Workbench projected tracked stereo imagery onto a table so specialists and observers could work around a shared virtual workspace. \timelinecite{KruegerFroehlich1994ResponsiveWorkbench}{73}}{1991}
    \eventcontinue{1995}{12.4}{\textbf{Japan:} Nintendo's Virtual Boy attempted mass-market stereoscopic play; its commercial failure documented the costs of poor comfort, limited presentation, and weak content support. \timelinecite{Strong1995VirtualBoy}{130}}{1991}
    \eventcontinue{1996}{15.5}{\textbf{Japan:} Sony's PLM-50 Glasstron packaged dual displays into a portable personal viewer, an important consumer precursor even though it was not a tracked VR system. \timelinecite{Sony1996Glasstron}{122}}{1991}
    \eventcontinue{2000}{18.6}{\textbf{United States:} WearTrack combined sourceless head orientation with head-relative hand tracking, explicitly targeting wearable computers and portable VR without site setup. \timelinecite{foxlin2000weartrack}{30}}{1991}
    \eventcontinue{2006}{21.7}{\textbf{Japan/global:} Nintendo's Wii Remote supplied inexpensive inertial sensing, infrared pointing, buttons, and mass-market availability that researchers repurposed for spatial interaction. \timelinecite{wingrave2010wiimote}{138}}{1991}
\end{tikzpicture}
\end{minipage}\timelinecoljoin
\begin{minipage}[t]{0.485\textwidth}
\vspace{0pt}\centering
\textbf{2007--2012: commodity sensing and mobile displays}\par\smallskip
\begin{tikzpicture}[x=1cm,y=-4.7mm,typnode/.append style={text width=5.9cm,font=\fontsize{6.5}{7.2}\selectfont}]
    \timelineaxis{24}
    \eventcontinue{2008}{0}{Commodity game controllers became a recognized 3D-interface design resource, with work on the Wiimote emphasizing both its convenience and its tracking limitations. {76, 138}}{2007}
    \eventcontinue{2010}{3.1}{\textbf{United States/Israel:} Microsoft Kinect brought camera-based, controller-free body tracking into homes at retail scale and quickly became reusable sensing infrastructure. {94, 126}}{2007}
    \eventcontinue{2011}{6.2}{\textbf{United States:} a smartphone-based HMD prototype placed the graphics system entirely onboard, demonstrating a lightweight wireless path using rapidly improving phone displays and processors. \timelinecite{olson2011design}{106}}{2007}
    \eventcontinue{2011}{9.3}{The UCAVE framework made portability, mobile communication, rapid setup, and ad hoc collaboration explicit VR system requirements. \timelinecite{DBLP:conf/cscw/BasuRJ12}{12}}{2007}
    \eventcontinue{2012}{12.4}{FOV2GO distributed a fold-flat cardboard smartphone viewer and development kit at IEEE VR, making a portable immersive prototype reproducible with the phone already in a user's pocket. \timelinecite{USCICT2012FOV2Go}{134}}{2007}
    \eventcontinue{2012}{15.5}{A wearable ``VR in the wild'' system integrated inertial tracking, obstacle sensing, and outdoor positioning into a self-contained mobile simulation platform. \timelinecite{DBLP:conf/vr/HodgsonBWBO12}{50}}{2007}
    \eventcontinue{2012}{18.6}{Immersive 3DUI on One Dollar a Day combined off-the-shelf components into an untethered sub-US\$900 system and quantified tracking performance. \timelinecite{6184191}{13}}{2007}
    \eventcontinue{2012}{21.7}{The Oculus Rift Kickstarter drew 9,522 backers and US\$2.4 million, converting a developer-kit HMD into a globally supported community platform. \timelinecite{Kickstarter2012OculusRift}{79}}{2007}
\end{tikzpicture}
\end{minipage}
\caption{Ubiquitous VR timeline: 1991--2012. Continue from the bottom of the left panel to the top of the right panel.}
\label{table:ubiquitousvrtimeline}
\end{table}

\begin{table}[p]
\centering
\footnotesize
\begin{minipage}[t]{0.485\textwidth}
\vspace{0pt}\centering
\textbf{2013--2015: smartphone and commodity platforms}\par\smallskip
\begin{tikzpicture}[x=1cm,y=-4.8mm,typnode/.append style={text width=5.9cm,font=\fontsize{6.5}{7.2}\selectfont}]
    \timelineaxis{24}
    \eventcontinue{2013}{0}{\textbf{United Kingdom:} Steed and Julier integrated an iPhone-based display, rendering, inertial sensing, and touch interaction into an immersive mobile VR system. \timelinecite{steed2013design}{125}}{2013}
    \eventcontinue{2013}{2.7}{The UCAVE pilot reported the evolution and usability of an experimental ubiquitous immersive interface assembled from off-the-shelf components. \timelinecite{6550213}{11}}{2013}
    \eventcontinue{2013}{5.4}{\textbf{United States:} Vance and colleagues examined Razer Hydra, Wiimote, and Kinect combinations, exposing the trade-offs involved in repurposing low-cost gaming inputs. \timelinecite{lu1virtual}{78}}{2013}
    \eventcontinue{2014}{8.1}{MuVR treated multi-user capacity, modularity, setup time, and external-infrastructure dependence as first-class deployment requirements. \timelinecite{DBLP:conf/vr/ThomasBGS14}{132}}{2013}
    \eventcontinue{2014}{10.8}{\textbf{France/United States:} Google Cardboard paired an open, low-cost viewer pattern with smartphones; by 2017 more than 10 million viewers had shipped. \timelinecite{Google2014Cardboard}{33}}{2013}
    \eventcontinue{2014}{13.5}{\textbf{South Korea/United States:} Samsung and Oculus introduced Gear VR for the Galaxy Note 4, coupling phone-scale compute and display with headset optics and dedicated sensors. \timelinecite{Samsung2014GearVR}{115}}{2013}
    \eventcontinue{2015}{16.2}{DSCVR showed how consumer displays and tracking could reduce the cost and specialization of a CAVE-like hybrid-reality environment. \timelinecite{ponto2015dscvr}{111}}{2013}
    \eventcontinue{2015}{18.9}{\textbf{South Korea/global:} Samsung announced the first consumer Gear VR at US\$99 with support for several flagship phones, moving phone-slot VR beyond an innovator edition. \timelinecite{Samsung2015ConsumerGearVR}{116}}{2013}
    \eventcontinue{2015}{21.6}{Consumer Rift, Vive, and PlayStation VR announcements consolidated a new platform era; their 2016 shipments are distinguished in the next panel. {87, 54, 123}}{2013}
\end{tikzpicture}
\end{minipage}\timelinecoljoin
\begin{minipage}[t]{0.485\textwidth}
\vspace{0pt}\centering
\textbf{2016--2018: packaged room scale and standalone VR}\par\smallskip
\begin{tikzpicture}[x=1cm,y=-4.5mm,typnode/.append style={text width=5.9cm,font=\fontsize{6.2}{6.9}\selectfont}]
    \timelineaxis{25}
    \eventcontinue{2016}{0}{\textbf{United States: 28 March.} Consumer Oculus Rift shipments turned the developer-kit revival into a supported PC VR platform. \timelinecite{Meta2016Rift}{87}}{2016}
    \eventcontinue{2016}{2.3}{\textbf{Taiwan/United States: 5 April.} HTC Vive packaged Lighthouse tracking and controllers for room-scale six-degree-of-freedom PC VR. \timelinecite{HTC2016Vive}{54}}{2016}
    \eventcontinue{2016}{4.6}{\textbf{Japan/global: October.} PlayStation VR connected immersive gaming to the installed PlayStation 4 base across Japan, North America, Europe, and Asia. \timelinecite{Sony2016PSVR}{123}}{2016}
    \eventcontinue{2016}{6.9}{A UCAVE study connected physical-fitness measures to performance and symptoms in an HMD-based maze, emphasizing user variability in deployable systems. \timelinecite{7460057}{9}}{2016}
    \eventcontinue{2017}{9.2}{Khronos announced the OpenXR working group, beginning a royalty-free application/runtime interface intended to reduce device fragmentation. \timelinecite{Khronos2017OpenXR}{63}}{2016}
    \eventcontinue{2017}{11.5}{\textbf{Germany:} ShareVR enabled an HMD user and non-HMD participants to share one co-located experience through projection, props, and asymmetric roles. \timelinecite{GugenheimerEtAl2017ShareVR}{40}}{2016}
    \eventcontinue{2017}{13.8}{\textbf{Taiwan/China:} HTC announced the standalone six-degree-of-freedom Vive Focus and Vive Wave, an open mobile platform aimed at China's fragmented headset market. \timelinecite{HTC2017ViveFocusWave}{55}}{2016}
    \eventcontinue{2018}{16.1}{\textbf{Germany:} FaceDisplay put cameras and touch screens outside an HMD so nearby people could see and manipulate the virtual scene. \timelinecite{GugenheimerEtAl2018FaceDisplay}{41}}{2016}
    \eventcontinue{2018}{18.4}{Oculus Go consolidated display, compute, tracking, audio, and software distribution into a self-contained US\$199 appliance. \timelinecite{Meta2018OculusGo}{83}}{2016}
    \eventcontinue{2018}{20.7}{\textbf{China/United States:} Lenovo Mirage Solo commercialized Google's WorldSense six-degree-of-freedom inside-out tracking in a standalone Daydream headset. \timelinecite{Google2018MirageSolo}{34}}{2016}
    \eventcontinue{2018}{23}{A UCAVE maze study compared audiovisual landmark designs, continuing the line's emphasis on navigation evidence rather than device novelty alone. \timelinecite{DBLP:journals/corr/abs-1805-09454}{10}}{2016}
\end{tikzpicture}
\end{minipage}
\caption{Ubiquitous VR timeline: 2013--2018. Continue from the bottom of the left panel to the top of the right panel.}
\end{table}

\begin{table}[p]
\centering
\footnotesize
\begin{minipage}[t]{0.485\textwidth}
\vspace{0pt}\centering
\textbf{2019--2021: inside-out tracking and deployable ecosystems}\par\smallskip
\begin{tikzpicture}[x=1cm,y=-4.5mm,typnode/.append style={text width=5.9cm,font=\fontsize{6.2}{6.9}\selectfont}]
    \timelineaxis{25}
    \eventcontinue{2019}{0}{\textbf{Finland: 19 February.} Varjo launched the professional VR-1 with a high-pixel-density focus display and integrated eye tracking for deployable industrial design and training. \timelinecite{Varjo2019VR1}{135}}{2019}
    \eventcontinue{2019}{2.3}{\textbf{Japan: 12 April.} Nintendo Labo VR combined Switch, cardboard Toy-Con, and family-oriented software into a simple buildable first VR experience. \timelinecite{Nintendo2019LaboVR}{105}}{2019}
    \eventcontinue{2019}{4.6}{\textbf{21 May.} Oculus Quest and Rift S shipped with camera-based Insight tracking; Quest delivered room-scale head and controller tracking without a PC or base stations. \timelinecite{Meta2019QuestRiftS}{84}}{2019}
    \eventcontinue{2019}{6.9}{\textbf{29 July.} OpenXR 1.0 established a stable, cross-vendor boundary between immersive applications and runtimes. \timelinecite{Khronos2019OpenXR10}{66}}{2019}
    \eventcontinue{2019}{9.2}{\textbf{December.} Quest hand tracking entered public beta, making controller-free input a software update for existing consumer hardware. \timelinecite{Meta2019QuestHandTracking}{85}}{2019}
    \eventcontinue{2020}{11.5}{Accessible by Design framed accessibility as an opportunity to redesign VR interaction, not a retrofit after mainstream deployment. \timelinecite{MottEtAl2020AccessibleByDesign}{99}}{2019}
    \eventcontinue{2020}{13.8}{The OpenXR Adopter Program and conformant Microsoft and Oculus runtimes turned the specification into a testable ecosystem commitment. \timelinecite{Khronos2020OpenXRConformance}{67}}{2019}
    \eventcontinue{2020}{16.1}{Quest 2 shipped at US\$299 with higher resolution and processing performance, accelerating standalone six-degree-of-freedom VR adoption. \timelinecite{Meta2020Quest2}{86}}{2019}
    \eventcontinue{2021}{18.4}{A CHI study showed that head-and-hand motion can identify users, making behavioral telemetry a deployment, security, and privacy concern. \timelinecite{LiebersEtAl2021BehavioralBiometrics}{77}}{2019}
    \eventcontinue{2021}{20.7}{\textbf{Taiwan/global:} Vive Focus 3 paired all-in-one inside-out tracking with enterprise manageability, replaceable power, and location-based deployment features. \timelinecite{HTC2021ViveFocus3}{56}}{2019}
    \eventcontinue{2021}{23}{Horizon Workrooms combined avatars, spatial audio, hand tracking, virtual screens, and mixed-reality desk features for remote collaboration. \timelinecite{Meta2021Workrooms}{88}}{2019}
\end{tikzpicture}
\end{minipage}\timelinecoljoin
\begin{minipage}[t]{0.485\textwidth}
\vspace{0pt}\centering
\textbf{2022--2023: global standalone competition and mixed reality}\par\smallskip
\begin{tikzpicture}[x=1cm,y=-4.5mm,typnode/.append style={text width=5.9cm,font=\fontsize{6.2}{6.9}\selectfont}]
    \timelineaxis{25}
    \eventcontinue{2022}{0}{\textbf{31 March.} W3C published a WebXR Candidate Recommendation Snapshot, standardizing browser access to immersive displays, sensors, and input. \timelinecite{W3C2022WebXR}{139}}{2022}
    \eventcontinue{2022}{2.3}{\textbf{China/Europe/Asia: 22 September.} PICO announced the all-in-one PICO 4 for Europe, with 2022 launches planned in Japan, South Korea, Singapore, and Malaysia. \timelinecite{PICO2022PICO4}{110}}{2022}
    \eventcontinue{2022}{4.6}{\textbf{25 October.} Quest Pro shipped with color stereoscopic passthrough, pancake optics, eye and face tracking, and self-tracking controllers. \timelinecite{Meta2022QuestPro}{89}}{2022}
    \eventcontinue{2022}{6.9}{\textbf{SIGGRAPH Asia: December.} UmeTrack integrated egocentric multi-view hand detection, tracking, and model fitting for robust articulated input. \timelinecite{HanEtAl2022UmeTrack}{42}}{2022}
    \eventcontinue{2023}{9.2}{\textbf{Japan/global: 22 February.} PlayStation VR2 launched with inside-out tracking, eye tracking, HDR display, headset feedback, and Sense controllers. \timelinecite{Sony2023PSVR2}{124}}{2022}
    \eventcontinue{2023}{11.5}{\textbf{5 June.} Apple announced Vision Pro and a high-resolution video-passthrough interaction model centered on eyes, hands, and voice. \timelinecite{Apple2023VisionPro}{4}}{2022}
    \eventcontinue{2023}{13.8}{An international accessibility review synthesized design barriers and interventions across VR, AR, and metaverse systems. \timelinecite{DudleyEtAl2023InclusiveImmersion}{26}}{2022}
    \eventcontinue{2023}{16.1}{Motion data from 55,541 users enabled highly accurate re-identification, demonstrating that ordinary VR trajectories can function as biometric data. \timelinecite{NairEtAl2023UniqueIdentification}{101}}{2022}
    \eventcontinue{2023}{18.4}{\textbf{France: October.} Lynx began R-1 shipments while committing to an open mixed-reality platform and accessible developer ecosystem. \timelinecite{Lynx2023R1}{80}}{2022}
    \eventcontinue{2023}{20.7}{\textbf{10 October.} Quest 3 shipped with color passthrough, a depth sensor, and room understanding in a mainstream standalone headset. \timelinecite{Meta2023Quest3}{90}}{2022}
\end{tikzpicture}
\end{minipage}
\caption{Ubiquitous VR timeline: 2019--2023. Continue from the bottom of the left panel to the top of the right panel.}
\end{table}

\begin{table}[p]
\centering
\footnotesize
\begin{minipage}[t]{0.485\textwidth}
\vspace{0pt}\centering
\textbf{2024: international spatial-computing deployment}\par\smallskip
\begin{tikzpicture}[x=1cm,y=-5mm,typnode/.append style={text width=5.9cm,font=\scriptsize}]
    \timelineaxis{22}
    \eventcontinue{2024}{0}{\textbf{2 February.} Apple Vision Pro shipped in the United States with eye-and-hand input, high-resolution passthrough, spatial video, and a desktop-class application model. \timelinecite{Apple2024VisionProShips}{6}}{2024}
    \eventcontinue{2024}{3.5}{A systematization of 74 publications organized VR data, threats, adversaries, and countermeasures, making privacy a first-class ubiquitous-system requirement. \timelinecite{MunillaGarridoEtAl2024PrivacySoK}{100}}{2024}
    \eventcontinue{2024}{7}{\textbf{15 April.} OpenXR 1.1 moved widely used extensions into the core, reducing fragmentation around local-floor spaces, foveated views, and palm poses. \timelinecite{Khronos2024OpenXR11}{68}}{2024}
    \eventcontinue{2024}{10.5}{\textbf{Asia/Europe: June--July.} Vision Pro expanded to China mainland, Hong Kong, Japan, Singapore, Australia, Canada, France, Germany, and the United Kingdom. \timelinecite{Apple2024VisionProInternational}{5}}{2024}
    \eventcontinue{2024}{14}{A controlled Vision Pro--Quest 3 comparison measured sub-centimeter relative tracking in most cases but much larger absolute errors in difficult conditions. \timelinecite{HuEtAl2024SpatialTracking}{57}}{2024}
    \eventcontinue{2024}{17.5}{\textbf{United States/South Korea: 12 December.} Google announced Android XR with Samsung and Qualcomm as a common platform for headsets and glasses. \timelinecite{Google2024AndroidXR}{35}}{2024}
\end{tikzpicture}
\end{minipage}\timelinecoljoin
\begin{minipage}[t]{0.485\textwidth}
\vspace{0pt}\centering
\textbf{2025--25 September 2026: platform convergence}\par\smallskip
\begin{tikzpicture}[x=1cm,y=-4.8mm,typnode/.append style={text width=5.9cm,font=\fontsize{6.5}{7.2}\selectfont}]
    \timelineaxis{23}
    \eventcontinue{2025}{0}{\textbf{20 May.} Google demonstrated Android XR glasses and began limited real-world testing, extending the platform from headsets toward everyday eyewear. \timelinecite{Google2025AndroidXRGlasses}{37}}{2025}
    \eventcontinue{2025}{2.8}{\textbf{17 September.} Meta announced Ray-Ban Display glasses and an EMG wristband; U.S. retail availability began 30 September. \timelinecite{Meta2025RayBanDisplay}{92}}{2025}
    \eventcontinue{2025}{5.6}{\textbf{15 October.} Apple announced an M5 Vision Pro revision and redesigned Dual Knit Band, an incremental compute and comfort update shipping 22 October. \timelinecite{Apple2025VisionProM5}{7}}{2025}
    \eventcontinue{2025}{8.4}{\textbf{South Korea/United States: 21 October.} Samsung introduced Galaxy XR, the first Android XR headset, supporting OpenXR, WebXR, Unity, and Android applications. {117, 36}}{2025}
    \eventcontinue{2026}{11.2}{\textbf{7 April.} Android XR added experimental 2D-to-3D spatialization, persistent app placement, physical-hand visibility, and tracking and accessibility improvements. \timelinecite{Google2026AndroidXRUpdate}{38}}{2025}
    \eventcontinue{2026}{14}{\textbf{19 May.} Google described audio-first and display Android XR eyewear. Audio glasses were slated for later autumn; no display-glasses date was given. \timelinecite{Google2026IntelligentEyewear}{39}}{2025}
    \eventcontinue{2026}{16.8}{\textbf{9 June.} W3C issued a new WebXR Candidate Recommendation Draft, continuing browser-based interoperability while the 2022 snapshot remained the latest CR Snapshot. \timelinecite{W3C2026WebXR}{140}}{2025}
    \eventcontinue{2026}{19.6}{\textbf{23 September.} Meta announced roughly 100~g split-compute VR glasses with a pocket puck, 5K micro-OLED display, passthrough, and eye/hand/voice input for spring 2027. \timelinecite{Meta2026VRGlasses}{93}}{2025}
\end{tikzpicture}
\end{minipage}
\caption{Ubiquitous VR timeline: 2024--25 September 2026. Continue from the bottom of the left panel to the top of the right panel. Announced products are distinguished from shipping products.}
\end{table}
\end{landscape}

\FloatBarrier

\subsection{Synthesis: five stages of ubiquity}

The milestones reveal five overlapping stages rather than a sequence of clean replacements:
\begin{enumerate}[leftmargin=*]
\item \textbf{Assembled access (1991--2006):} researchers and public venues combined specialized and consumer components to lower the laboratory barrier.
\item \textbf{Commodity sensing and mobile displays (2007--2012):} game controllers, depth cameras, smartphones, and inexpensive viewers became reusable building blocks.
\item \textbf{Packaged consumer platforms (2013--2018):} phone-slot viewers, PC headsets, room-scale tracking, and early standalone systems converted prototypes into supported products.
\item \textbf{Standalone ecosystems (2019--2023):} inside-out tracking, hand sensing, stores, interoperability, and international competition reduced setup while expanding platform dependence.
\item \textbf{Spatial-platform convergence (2024--2026):} passthrough mixed reality, eye-and-hand input, learned computation, and AI services brought VR, AR, and everyday computing into a shared platform layer.
\end{enumerate}
Across these stages, visible equipment decreases while responsibility shifts toward perception software, runtimes, cloud services, governance, and long-term ecosystem support. This transition motivates the four contemporary trajectories considered next.

\section{Current trajectory: integrated spatial computing}
\label{sec:tcurrenttrendsinvr}

The smartphone and app-store era was a pivotal stage in VR democratization, but the trajectory through the cutoff date of this article is more accurately described as \textit{integrated spatial computing}. Standalone and mixed-reality headsets combine display, compute, visual--inertial tracking, hand and eye input, environment sensing, and network services. This reduces visible external infrastructure and setup time, but it also transfers more responsibility to onboard perception software and platform governance.

\subsection{Integrated sensing and mixed reality}
Inside-out tracking has replaced external tracking infrastructure for many uses, and optical hand tracking increasingly supplements or replaces controllers \citep{HanEtAl2022UmeTrack}. Passthrough video and spatial mapping allow virtual objects to share a coordinate frame with the physical environment. The resulting system is easier to deploy, but its quality depends on calibration, lighting, occlusion handling, map persistence, and the stability of spatial anchors. Comparative measurements of contemporary headsets show that tracking performance remains a systems question rather than a solved feature checkbox \citep{HuEtAl2024SpatialTracking}.

\subsection{Perceptual and learned computation}
Rendering is increasingly organized around what a user can see and attend to. Foveated rendering concentrates work near the gaze point, while attention-aware methods recognize that gaze and visual importance are related but not identical \citep{AlbertEtAl2017FoveatedRendering,KrajancichEtAl2023AttentionAware}. Learned scene representations such as neural radiance fields and 3D Gaussian splatting also shift scene capture and view synthesis toward data-driven representations \citep{MildenhallEtAl2020NeRF,KerblEtAl2023GaussianSplatting}. These methods can improve fidelity or efficiency, but they introduce new requirements for capture provenance, update latency, failure detection, and predictable behavior.

\subsection{Interoperability and distribution}
OpenXR provides a common application-facing API across conformant XR runtimes, while WebXR exposes immersive capabilities through the web platform \citep{Khronos2024OpenXR11,W3C2026WebXR}. Standards do not remove all fragmentation---input profiles, extensions, performance budgets, stores, and platform policies still differ---but they make portability a design objective rather than a later porting exercise. For ubiquitous VR, this layer is as consequential as miniaturized hardware because it affects who can create, deploy, preserve, and revisit an experience.

\subsection{Human constraints and trust}
More sensors and features do not automatically produce a better experience. Comfort and safety depend on content motion, locomotion, exposure, fit, sensory coherence, and individual variability as well as display specifications \citep{SaredakisEtAl2020VRSickness}. Accessibility is similarly architectural: inclusive input alternatives, captions and spatial-audio substitution, reach and mobility accommodations, adjustable presentation, and recoverable interaction should be considered from the beginning \citep{MottEtAl2020AccessibleByDesign,DudleyEtAl2023InclusiveImmersion}. Finally, head, hand, eye, voice, and spatial data may support highly identifying or sensitive inferences \citep{NairEtAl2023UniqueIdentification,MunillaGarridoEtAl2024PrivacySoK}. A ubiquitous system must therefore minimize and protect data while making sensing behavior intelligible to users.

\section{Open research agenda: from access to accountable deployment}

The chronology suggests five priorities for work beyond hardware miniaturization.

\subsection{Reliability outside controlled spaces}
Systems should characterize tracking confidence, degrade safely under poor lighting or occlusion, and recover from lost maps or anchors without forcing a complete restart. Evaluation should report not only mean accuracy but also failure frequency, recovery time, environmental conditions, and consequences for the task.

\subsection{Perceptual efficiency with accountable adaptation}
Gaze-contingent rendering, learned reconstruction, and AI-mediated interfaces should expose uncertainty and failure modes. Research should test whether optimizations perform consistently across visual ability, physiology, task, and device fit rather than treating a single average observer as universal.

\subsection{Accessibility as a platform property}
Applications need interoperable accessibility semantics and input substitution, not patches confined to individual applications. Authors should involve disabled participants early, document supported interaction modes, and make locomotion, reach, audio, text, contrast, and timing configurable \citep{MottEtAl2020AccessibleByDesign,DudleyEtAl2023InclusiveImmersion}.

\subsection{Privacy-preserving spatial computing}
Research should reduce the collection of raw motion, eye, voice, bystander, and environment data; favor on-device processing where appropriate; and develop meaningful permission models for derived inferences rather than sensor access alone. Evaluation should include re-identification and inference risk as system metrics \citep{NairEtAl2023UniqueIdentification,MunillaGarridoEtAl2024PrivacySoK}.

\subsection{Evidence beyond novelty}
Longitudinal and comparative studies are needed to establish when immersive delivery improves learning, health, work, or collaboration and when a less immersive medium is preferable. Reproducible reporting should include the device and runtime, software version, tracking mode, frame timing, locomotion method, exposure, adverse-event handling, accessibility accommodations, and relevant user characteristics.

\section{Conclusion}

The history of VR is not a simple progression toward smaller headsets or more realistic images. Its recurring advance is integration: displays, sensing, computation, interaction, and content become coordinated well enough to close the perception--action loop for a useful task. Smartphones accelerated access to compact displays and inertial sensing; standalone headsets extended that integration to positional tracking, hand and eye input, mixed reality, and networked platforms.

That integration also changes the meaning of ubiquitous VR. External equipment may disappear while sensing, platform dependence, and data collection become more extensive. Portability and price remain important, but an experience is not broadly deployable if it is fragile, incompatible, inaccessible, unsafe, or opaque about the information it captures. The central design challenge after 2026 is therefore not only to make VR more immersive, but to make immersive systems dependable, inclusive, interoperable, and worthy of trust.

\section*{Acknowledgments}
This article grew from my Ph.D. dissertation research under the supervision of Dr. Kyle Johnsen at the University of Georgia.

%----------------------------------------------------------------------------------------
%	REFERENCE LIST
%----------------------------------------------------------------------------------------

\begingroup
\small
% Register all references used by the plain-text timeline citations.  Keeping
% this metadata outside TikZ and landscape environments prevents arXiv's HTML
% conversion from attaching citation anchors to transformed node coordinates.
\nocite{
  CummingsBailenson2016Presence,Meta2016Rift,Microsoft2016HoloLensShips,HTC2016Vive,
  OrtsEscolanoEtAl2016Holoportation,Meta2016Touch,Khronos2017OpenXR,GugenheimerEtAl2017ShareVR,
  Google2017StandaloneDaydream,KotaruKatti2017WiCapture,AlbertEtAl2017FoveatedRendering,GugenheimerEtAl2018FaceDisplay,
  Meta2018OculusGo,Google2018MirageSolo,RomeroRondonEtAl2018FoveatedStreaming,Khronos2018OpenXRDemos,
  Microsoft2019HoloLens2,Khronos2019OpenXR090,Meta2019QuestRiftS,Khronos2019OpenXR10,
  KaplanyanEtAl2019DeepFovea,Meta2019QuestHandTracking,RadiantiEtAl2020HigherEducation,Khronos2020OpenXRConformance,
  MildenhallEtAl2020NeRF,Meta2020Quest2,Microsoft2021Mesh,LiebersEtAl2021BehavioralBiometrics,
  Meta2021Workrooms,W3C2022WebXR,Meta2022QuestPro,HanEtAl2022UmeTrack,
  Apple2023VisionPro,KerblEtAl2023GaussianSplatting,KrajancichEtAl2023AttentionAware,NairEtAl2023UniqueIdentification,
  Meta2023Quest3,Apple2024VisionProShips,MunillaGarridoEtAl2024PrivacySoK,Khronos2024OpenXR11,
  Meta2024Orion,HuEtAl2024SpatialTracking,Google2024AndroidXR,Google2025AndroidXRGlasses,
  Meta2025RayBanDisplay,Apple2025VisionProM5,Samsung2025GalaxyXR,Google2025GalaxyXR,
  Google2026AndroidXRUpdate,Google2026IntelligentEyewear,W3C2026WebXR,Meta2026VRGlasses,
  Pausch:1991:VRF:108844.108913,HCIMuseum1991Virtuality,pausch1993user,KruegerFroehlich1994ResponsiveWorkbench,
  Strong1995VirtualBoy,Sony1996Glasstron,foxlin2000weartrack,wingrave2010wiimote,
  lee2008hacking,Microsoft2010Kinect,suma2011faast,olson2011design,
  DBLP:conf/cscw/BasuRJ12,USCICT2012FOV2Go,DBLP:conf/vr/HodgsonBWBO12,6184191,
  Kickstarter2012OculusRift,steed2013design,6550213,lu1virtual,
  DBLP:conf/vr/ThomasBGS14,Google2014Cardboard,Samsung2014GearVR,ponto2015dscvr,
  Samsung2015ConsumerGearVR,Sony2016PSVR,7460057,HTC2017ViveFocusWave,
  DBLP:journals/corr/abs-1805-09454,Varjo2019VR1,Nintendo2019LaboVR,MottEtAl2020AccessibleByDesign,
  HTC2021ViveFocus3,PICO2022PICO4,Sony2023PSVR2,DudleyEtAl2023InclusiveImmersion,
  Lynx2023R1,Apple2024VisionProInternational}
\bibliographystyle{acm}
\bibliography{references}
\endgroup

%----------------------------------------------------------------------------------------

\end{document}